# Label free sub-diffraction imaging using non-linear photon avalanche backlight

S.Karmegam[1], M.Szalkowski[2], M.Misiak[1], K.Prorok[1], D. Szymański[1], A.Bednarkiewicz[1*]

1. Institute of Low Temperature and Structure Research, Polish Academy of Sciences, ul. Okólna 2, Wroclaw 50-422, Poland

2. Nanophotonics Group, Institute of Physics, Faculty of Physics, Astronomy and Informatics, Nicolaus Copernicus University in Toruń, ul. Grudziądzka 5, 87-100 Torun, Poland

*E-mail: a.bednarkiewicz@intibs.pl

**ABSTRACT**

Optical imaging below the limit of light diffraction offers an unprecedented opportunity to study outlook, organization, interactions or *in-situ* functioning of sub-micrometer, highly transparent objects such as subcellular structures *in vitro*, thin layers or nano-engineered devices. However, most of current methodologies require to use specially designed luminescent labels, which not only may affect the properties of the sample itself, but often are (photo)toxic, susceptible to photobleaching, offer limited color combinations or specificity of labeling. Moreover, the dedicated fluorescence based super-resolution optical techniques are often technically complex and cumbersome to use. From the other hand, the existing non-destructive, non-invasive and label-free super-resolution imaging (SRI) methods are also challenging, complex and elusive to apply. To address these issues, here we propose and experimentally demonstrate a new concept of label-free sub-diffraction optical imaging. The transmission avalanche backlight (TAB) microscopy exploits huge optical non-linearities of photon avalanching materials, which are acting as a virtual near-field nano aperture - a diffraction limited backlight of the actual sample. Such approach enables to augment imaging contrast of highly transparent samples and thin layers, by translating small attenuation and scattering loses occurring on these translucent samples into amplified modulation of luminescence intensity of the avalanche backlighted substrate (ABS). Simultaneously, at no additional cost, sub-diffraction imaging is achieved with simple, single beam laser scanning microscopy setup, leading to ca. 70 nm optical resolution. This far-field, label-free, raster scanning imaging technique, with augmented contrast and optical imaging resolution below diffraction limit, may become pivotal for studies in biology, physics, materials science, nanophotonics and nanoengineering.

**INTRODUCTION**

Current progress and understanding in (micro)biology [1–3], physiology [4–6], physics [7] or technology [8] are often possible thanks to the ability to visualize the morphology, appearance, functioning and behavior of submicron or nanometer objects. These advancements became possible thanks to developing groundbreaking concepts of optical super-resolution imaging (SRI), together with dedicated instruments and sophisticated data acquisition or analysis (**Figure S1**) [9–13]. However, most of these techniques require fluorescent labeling [14–17] of the delicate samples, which has driven designing special fluorescent labels enabling bio-specific interaction with target sites, limited (photo)toxicity as well as appropriate fluorescent and chemical properties[18]. Nevertheless, to address the issues related to photobleaching, phototoxicity, spectral bleed-through or limited multiplexing capability of fluorescent molecules, new classes of luminescent nanoparticle-based labels have been also developed. Favorably, the latter have shown enhanced photostability and blinking-free operation, as well as background free detection, which owes to efficient upconversion of near infrared (NIR) excitation to NIR/visible spectral range[19,20]. Despite numerous existing challenges related to bio-functionalization, bio-specificity, multiplexing capabilities or sensitivity of such fluorescent label-based SRI methods have been successfully addressed, incorporation of exogenous labels may affect the delicate machinery and activity of living cells or influence the inherent properties of photonic devices under investigation. In consequence, a fluorescent snapshot of the perturbed system is achieved, rather than prolonged, remote and damage-free observation of native sample. On the top of that, the SRI optical techniques are often complex and cumbersome to apply due to sophisticated design of optical instruments [21–23] or tedious fluorescence images collection, analysis and interpretation [11,24,25]. Although new SRI techniques have been proposed, such as plasmonic-SIM [26,27], hyperbolic metamaterials-SIM [28,29], solid immersion lens approach [30,31], which however do not solve all constrains of existing techniques. Compromising remote readout, SRI may be also achieved with atomic force microscopy (AFM) or near-field techniques, such as scanning near-field optical microscopy (SNOM) or tip-enhanced near-field optical microscopy (TENOM), whose applications are however restricted to those nanoobjects, which allow for direct contact of the tip with their surface. Moreover, numerous technical difficulties, such as AFM/SNOM/TENOM probes contamination and image quality have so far hindered wider adoption of these methods in biology. Therefore, novel, label-free, non-destructive, technically simple and direct (reconstruction-free) imaging techniques for *in-situ* visualization of highly transparent samples below diffraction limit of light remain of a great interest, but also a great challenge.

To circumvent the alteration of the properties or dynamics of the biological or photonic samples by incorporation of exogenous fluorescent labels, and still preserve an access to sub-diffraction optical imaging, a few label-free techniques have been proposed[32,33]. The Abbe–Rayleigh diffraction limit can be overcame by generating super oscillatory wavefront [34,35], using of super oscillatory lenses (SOL) and SOL microscopes [36,37], coherent scattering under oblique angle illumination [38], phase SIM [39], expansion stimulated Raman vibrational imaging [40], photon-reassignment [41] or reconstruction topological microscopy [36]. Also STED approach, one of the current gold standards in fluorescence microscopy, can be combined with label-free saturable absorption of the sample [42]. Other methods exploit photothermal gradient effects in the sample [43], graphene quenched backlight $Pr^{3+}$ fluorescence [44] or pump-probe imaging of endogenous chromophores [45]. Moreover, femtosecond stimulated Raman spectroscopy has been engaged for label-free identification of vibrational fingerprints of chemicals or functional groups [46] below the diffraction limit. Despite achieving factual resolution enhancement without fluorescent labels [32,33], great majority of these methods suffer from one or more drawbacks, such as complexity, cost inefficiency, near-field readout, computationally heavy image reconstruction with a priori knowledge on light pattern or the objects themselves, the requirement of sophisticated on-demand designed elements (lenses, metamaterials, plasmonic layers, etc.), thus can often hardly be implemented for observations of biological or highly transparent samples. Solution to these obstacles can be found in radically novel imaging concepts and unconventional materials, such as highly nonlinear photon avalanching (PA) crystals.

The PA phenomenon was discovered in 1979 in bulk $LaCl_3$ $Pr^{3+}$ doped single crystals [47] and adopted to count medium infrared (MIR) photons. For over 40 years, a few studies demonstrated PA in other lanthanides (Ho, Tm, Sm, Nd, Er) doped bulk materials (single crystals, waveguides, glasses) mostly in cryogenic temperatures (4.2-77 K), exclusively for MIR photons detection or upconversion lasers[48–50]. The PA relays on positive loop of sequential excited state absorption (ESA) and energy cross-relaxation (CR) between neighbor lanthanide ions (**Figure 1a**), which, above distinct pump power threshold, exhibit very steep relationship between pump ($I_P$) power and luminescence ($I_L$) following power law, $I_L = (I_P)^S$, where the non-linearity index (*S*) can reach values much beyond 10, 20-40 [51–54] on average, up to even 500 [55] (**Figure 1b**). Only recently PA has been demonstrated in nano- and micro-sized inorganic materials co-doped with selected lanthanide ions [51,52,54]. Due to their highly non-linear behavior and complex kinetic photoluminescence, these materials have proven to be suitable for facile PA single beam sub-diffraction imaging (PASSI) [51,52,54], photo-switching and sub-nanometer localization imaging[56], as well as reservoir computing[57], pico-newton force[58] and temperature sensing [59]. In particular, direct sub-diffraction resolution (*λ*/5, ca.70 nm for *S*=23) imaging was proposed [60] and shown [51] using highly non-linear 800 nm luminescence ($I_L$) of thulium ($Tm^{3+}$) doped inorganic photon avalanching nanoparticles (ANPs) under 1059 nm photoexcitation wavelength. Favorably, as a result of these highly nonlinear nano-labels, the ANP based SRI relays on a technically simple and affordable confocal laser scanning microscopes subject to minor modifications – such as implementation of a customized filter cube and dedicated NIR laser. However, these novel possibilities do not refrain from using (biofunctionalized) avalanching nanoparticle labels.

In opposite to PASSI or the other conventional fluorescent-label based imaging method, here we propose, simulate and experimentally verify a radically different SRI concept of transmission avalanche backlight (TAB) microscopy (**Figure.1**). TAB exploits the same, highly non-linear luminescent properties of avalanching materials (**Figure.1c**), which however this time originate from a bulky PA crystal acting as a large area homogenous avalanche backlight substrate (ABS) which is backlighting the pristine non-labeled sample. When the nonlinear ABS crystal is photoexcited with a diffraction-limited excitation beam, the spot it reaches mimics a sub-diffraction, near-field virtual aperture (VA). Concurrently, tiny fluctuations of the excitation intensities just beneath the very confined area of the semi-transparent sample are translated by the non-linear photon avalanche phenomenon to strong variations of photoluminescence, and thus augment imaging contrast (**Figure 1d**). Because this virtual aperture is smaller than the diffraction limited excitation spot, it enables sub-diffraction optical resolution, similarly to SNOM method (**Table S1**). However, in opposite to SNOM, the physical aperture of the scanning tip moving above the surface of the sample is replaced with a virtual aperture that is scanning the sample throughout its volume along the optical path of the laser. With no labels, the reading is performed solely due to interaction (i.e., attenuation, scattering or energy transfer) of the focused excitation beam with the nano-structured and/or highly transparent sample (**Figure 1d-e**). Although this interaction occurs between the sample and ABS substrate in near-field, the TAB image is generated and read indirectly in far-field fluorescence mode. For the reasons given above, TAB offers improved contrast and optical resolution without the need for luminescent nanolabels (**Figure.1g-j**)

In the newly proposed TAB microscopy concept, instead of measuring the incident and transmitted photon flux, light transmitted through the specimen is used to photoexcite luminescence of the highly non-linear photon avalanching backlight substrate (**Figure 1c**). In consequence, highly (but not perfectly) translucent sample (composed of single cells, nanostructures, thin layers, etc.) reduces the pumping intensity that reaches ABS (**Figure 1e** vs **d**). The effective excitation intensity (**Figure 1f**, red solid arrow) is thus slightly lower than the pristine one (red hashed arrow), which, owing to the non-linearity of PA (**Figure 1b**), is then translated to significantly weaker PA emission intensity (**Figure 1f**, blue solid arrow) than for bare ABS (**Figure 1f**, blue hashed arrow). Therefore, without labeling the sample itself, the luminescence intensity of the PA substrate responds to the presence of the physical obstacles located on the paths and allows to quantitate transmittance thereof at the pumping wavelength. Quantitative evaluation can be performed using parameters defined in Methods section, such as luminescence intensity ratio (LIR) (**Eq.1**), contrast (C) (**Eq.2**), contrast potential ($\pi_x$) (**Eq.3**) or $\omega_{95\%}$ (**Figure 1h-k**). For example, presence of the sample showing 95% transmission reduces the PA emission intensity down to 36% of the original intensity for ABS with nonlinearity equal 20 (**Figure 1h** and **j**). This in turn improves the contrast and enables visualization of such highly transparent heterogenous samples with greater details. Moreover, the simulations indicate that the edge of the thin layers can be visualized with higher optical resolution, following $S^{-0.5}$ rule of thumb [60], down to 100 nm for nonlinearities S = 30.

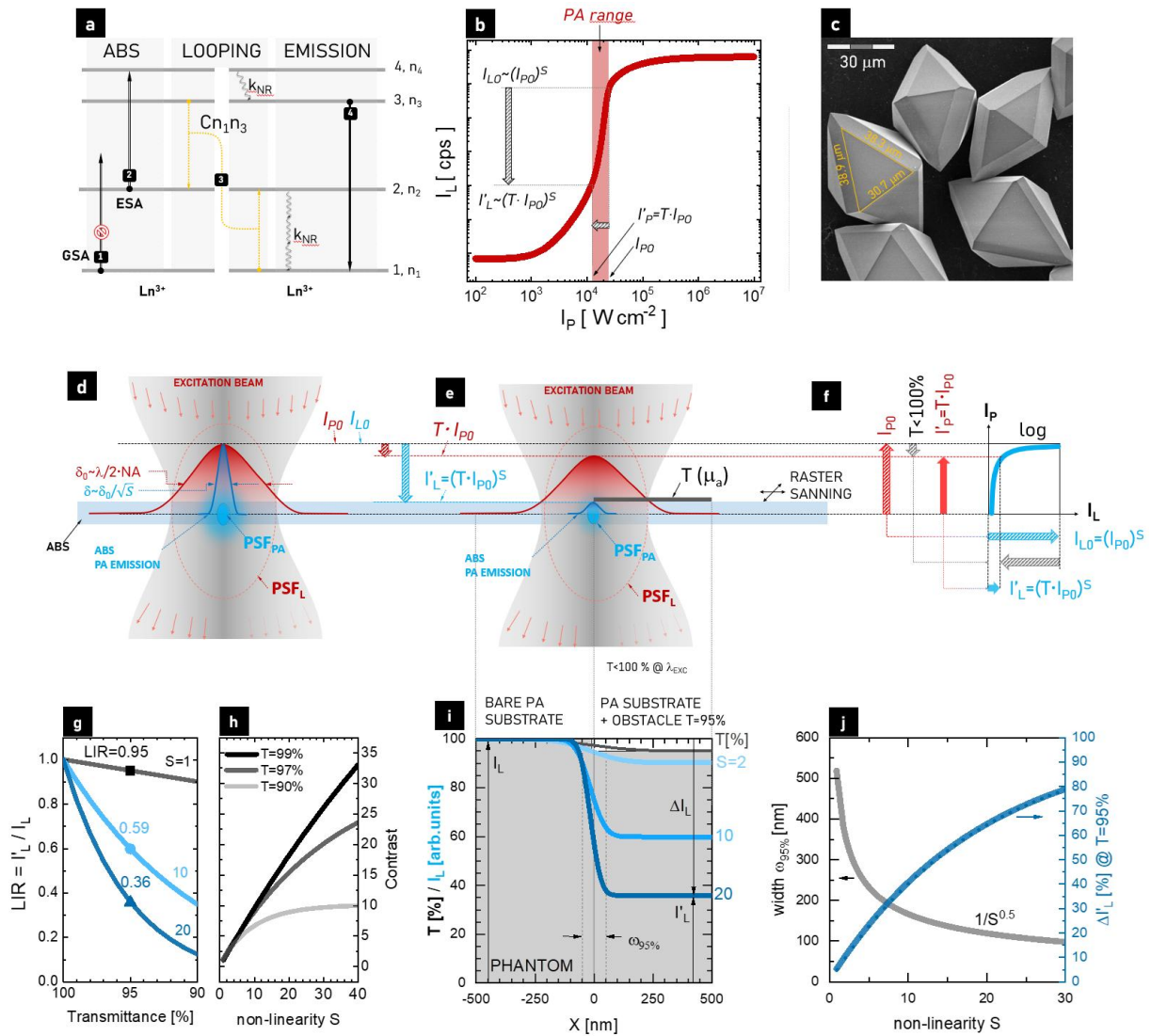

**Figure 1. The principles behind Photon Avalanche (PA) and Transmission Avalanche Backlight microscopy (TAB).** (a) General mechanisms leading to photon avalanche emission – absent GSA, efficient ESA resonant absorption and energy looping leads to (b) highly non-linear emission intensity dependence versus pump power intensity (PPD); (c) representative SEM image of photon avalanching 3%Tm$^{3+}$ doped LiYF$_4$ microcrystals, the foot forms a triangle with the average area of ca. 540 μm$^2$. General mechanism behind TAB microscopy – (d) diffraction limited laser excitation beam hits avalanche backlight substrate (ABS) with pump intensity $I_{P0}$, owing to its high non-linearity index ($S$), the effective point spread function (PSF$_{PA}$) is reduced by a factor of $\sqrt{S}$ as compared to a linear case (PSF$_L$); (e) a semi-translucent obstacle (T < 100%) reduces excitation beam intensity to $T \cdot I_{P0}$, and this tiny drop is translated to huge PA emission intensity drop on the non-linear PA substrate by a factor of $T^S$ from original $I_{L0} = (I_{P0})^S$ to $I'_{L0} = (T \cdot I_{P0})^S$. (f) Explanation of how the photoexcitation (red arrows) upon presence of obstacle (grey arrow) is converted to luminescence (blue arrows) by the non-

linear photon avalanche backlight substrate; the dashed and solid arrows indicate original (without obstacle) and effective (with obstacle) signals. (g) Simulation of PA luminescence intensity across the obstacle of given transmittance for a linear (S=1, grey) and non-linear (S=10 - blue and 20 - dark blue); (h) the contrast enhancement for various sample transmissions and rising PA non-linearity index; (i) a PA intensity cross section across the case presented on panel (e) for the non-linearity index S=2,10,20; (j) the simulations of non-linearity *S* dependent width $\omega_{95\%}$ (left) and PA luminescence loss (right); the $\omega_{95\%}$ and $\Delta I'_L$ are defined graphically on panel (i); the $\omega_{95\%}$ indicate the width (in nanometers) corresponding the intensity loss from 5% to 95 % at the semi-translucent layer edge; the $\Delta I'_L$ is the PA intensity loss due to sample of 95% transmission as compared to pristine PA substrate emission intensity.

The rationale behind augmented optical resolution is less straightforward to be understood and quantified with a mathematical formula, despite it originates from exactly the same effects as the one standing behind PASSI resolution enhancement [60]. When the excitation diffraction limited beam enters the area with an obstacle (**Figure 1e**) of decreased transparency T, the non-linear photon avalanche emission drops dramatically in response to the displacement from non-attenuated area. The displacing of the diffraction limited pump beam spot away from non-attenuated region towards the semi-transparent area decrease as $T^S$, unlike one finds in a linear backlighted case, where the excitation pump power (and thus fluorescence intensity) drops linearly ($\sim T^1$) with the transmittance decrease. This means the effective point spread function (PSF) volume in non-linear avalanching materials changes with an inverse square root of the non-linearity index S (**Figure 1j**). To further clarify and visualize these new possibilities, we have employed the same algorithms (described shortly in TAB simulations section) as we used to introduce the concept on sub-diffraction imaging with non-linear labels[51,60]. We have simulated the response of the ABS substrate to attenuating layers and particles for various non-linearities (**Figure 2**, **TAB simulations** chapter).

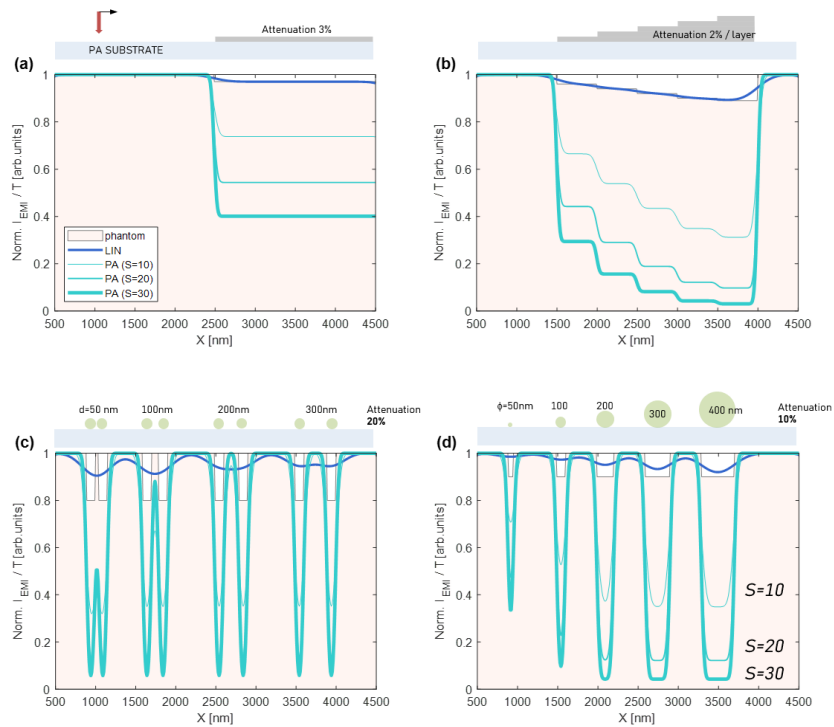

**Figure 2. Simulations of TAB imaging with various samples types and rising non-linearity index.** Cross-section of the TAB image across the (a) thin layer with 3% attenuation, (b) multiple thin overlapping layers with attenuation of 2% each, (c) two nanoparticles with attenuation of 20% at various distances

(d=50, 100, 200, 300 nm), (d) nanoparticles with attenuation of 10% and rising size (ϕ=50, 100, 200, 300, 400 nm). Blue line – image cross section accounting for linear attenuation, green lines – image cross-sections accounting for avalanching substrate with rising nonlinearity index (S=10-thin, 20-medium, 30-thick line).

Clearly, non-linearity index dependent contrast enhancement and augmented optical resolution can be observed in the PA intensity cross-sections for thin layer of various attenuation (Figure 2a) or thickness (Figure 2b), nanoparticles ensembles (Figure 2c) or nanoparticles of different size (Figure 2d). For the sake of simplicity, in this modelling we assumed perfectly homogenous avalanching backlighting of the sample and we disregarded light scattering, multiple reflections or light diffraction effects on the sample. Nevertheless, these phenomena in real experimental configuration may impact the interpretation of the TAB images.

To validate the TAB concept, we have first synthesized bulk ABS crystal, which was doped with $Tm^{3+}$ ions at increased concentrations to purposefully augment energy cross-relaxation and energy looping to achieve photon avalanching properties. The $Tm^{3+}$ doped $LiYF_4$ crystals were made using hydrothermal synthesis method (*Synthesis of 3%Tm LiYF₄ microcrystals* chapter). These crystals demonstrated well-defined reproducible morphology and relatively large size, although some crystal-to-crystal size (chapter S2 of SI) and negligible PA properties variations within single ABS crystal could be observed (chapter S3 of SI). Moreover, these ABS crystals exhibited sufficient non-linearity index of 8.5±1.0 and mild 11-16 $kW/cm^2$ PA thresholds (Figure S3.1). As we demonstrated, these ABS-es were also resistant to photodarkening effects that was found in $NaYF_4$ nanoparticles[56] – no intensity decrease upon strong (>550 $kW/cm^2$) and prolonged (120 s) illumination with a focused 1064 nm laser (Figure S3.2) was found. Moreover, some acceptable variation of intensity was noticed (Figure S3.3), which can be ascribed to small inhomogeneities in local concentration of $Tm^{3+}$ avalanching and Y/F structural ions (Figure S2.2 and Table S2.1). We noticed increased PA signal at the edges and corners of the ABS originating from enhanced excitation or emission light scattering. While we did not see other optical effects like multiple reflections or light diffraction in our $Tm^{3+}$ doped $LiYF_4$ ABS, some of these phenomena we found in other $Tm^{3+}$ doped β-$NaYF_4$ ABS crystals that were preliminarily evaluated (Figure S3.4). In the latter crystals, we did observe most probably light diffraction or reflection from the flat edges of the fluoride microcrystals surrounded by glass or air, which owes to the fact the excitation wavelength and beam size is comparable or smaller than the ABS size itself. Such effects indicate the importance of consideration of these phenomena for future studies and developing optimized ABS substrates.

Beside optical effects, we expect that because of the high non-linearity, the PA luminescence may vary strongly depending on the perturbations related to interaction between the ABS substrate and the sample[61], as well as the quality (composition homogeneity, flatness, lack of cracks, photostability) of the ABS crystals themselves. The natural consequence of label-free TAB imaging is lack of specificity, e.g. the nano- micro-structures are visualized because they are capable to attenuate the excitation beam. Because the attenuation is rather ubiquitous and not very specific (like chemical/biochemical characteristics suitable for fluorescent tagging with exogenous fluorescent molecules or Raman signal), pure attenuation would limit the interpretation of TAB images to the visually enhanced morphology of the samples. However, some materials, e.g. noble metal nanostructures, graphene, organic compounds, fluorophores, etc., can, beside light attenuation, specifically interact with the excitation beam and with ABS substrate. These interactions span from quenching to enhancement. To understand possible scenarios, we evaluated avalanching microcrystals (*Synthesis of 3%Tm LiYF4 microcrystals* chapter) and employed differential rate equation model (*Modeling of PA properties* section) to theoretically describe, model and predict the impact of various phenomena on the susceptibility of the ABS crystals to external and internal factors such as homogeneity of Tm3+ doping, attenuation and quenching. To gain this information, one would need to perform locally, pump power dependent luminescence intensity characterization and this is technically simple and affordable to be included during TAB imaging. For most simple - plain attenuation case (*Figure 3a*), excitation light beam is attenuated and no interaction between ABS and the sample occurs. In such a case the pump power dependent relationship is apparently shifted to larger PA thresholds (**Figure 3a**), but the "S"-shape curve shape should be preserved. However, due to non-trivial operation of the system featuring increased non-linearity, the contrast enhancement (**Eq.1**) and luminescence intensity ratio (LIR) (**Eq.1**) for the corresponding pump power dependent profiles (2nd and 3rd rows of **Figure 3**) reveal non-monotonic behavior. The best contrast improvement can be obtained at pumping power fixed to value at which the avalanching regime changes to saturation ($I_{P0}$ in **Figure 1b**, **Figure 3** vertical dashed line). The next cases worth considerations are the quenching of either emitting (**Figure 3b**) or looping (**Figure 3c**) level as a result of non-radiative energy transfer from the excited ABS to the sample. Quenching will typically occur for very short distances below 10 nm, and should be observed for plasmonic or wide bandgap materials. There are some distinct differences in those profiles and between them and the attenuation profiles. The attenuation induced *LIR($I_P$)* profile is more symmetric around central peak than the other two. It is straightforward to distinguish between two variants of quenching, because the quenching of the looping level not only reduce its intensity but also the slope in response to growing quenching rate. Nevertheless, the attenuation and looping level quenching profiles show many similarities. Therefore, unequivocal interpretation of the ABS-sample interaction mechanism, could require a control experiment. Assuming perfect ABS-to-ABS homogeneity of the PA properties, one could replicate the TAB image of the sample performed with the pristine ABS and compare it to experiment with some spacing (e.g. with a polymer or undoped ABS shell etc.) between the sample and ABS crystal. While the first map would show complex behavior, the reference TAB map would exclude non-radiative interactions and keep attenuation phenomenon only. In consequence, such additional experiment would enable differentiation to extract pure interaction component from pure attenuation.

Beside effects described above, in large size ABS crystals, some inhomogeneities may occur in their composition, which will affect the CR efficiency and energy looping (**Figure S3.3**). This in turn will ultimately change the local behavior of the ABS crystal and will impact the interpretation and quantification of the observed TAB maps. To some extent, this can be observed in experimental TAB image (**Figure S3.3a**). While the SEM-EDX maps of the ABS confirm the mostly homogenous distribution of 3% Tm on the surface of ABS, local variations of Y and F atoms distributions have been found, and thus validate our interpretation (**Figure S2.2, Table S2.1**). Numerical modeling indicate that decreasing or increasing ($\Delta k_{CR}$ = -10, -5, +5, +10% ) of cross-relaxation rate, $k_{CR}$, is capable to modify the PA behaviour, what ultimately affects the LIR ratio (**Figure S3.3d-e**). Therefore, either further optimization of ABS crystals is required, or care must be taken to properly interpret local variations of the TAB maps. Nevertheless, the experimentally observed variations in our ABS crystals were spread over larger areas and not tightly localized, thus we found them to provide sufficient quality to perform further TAB imaging.

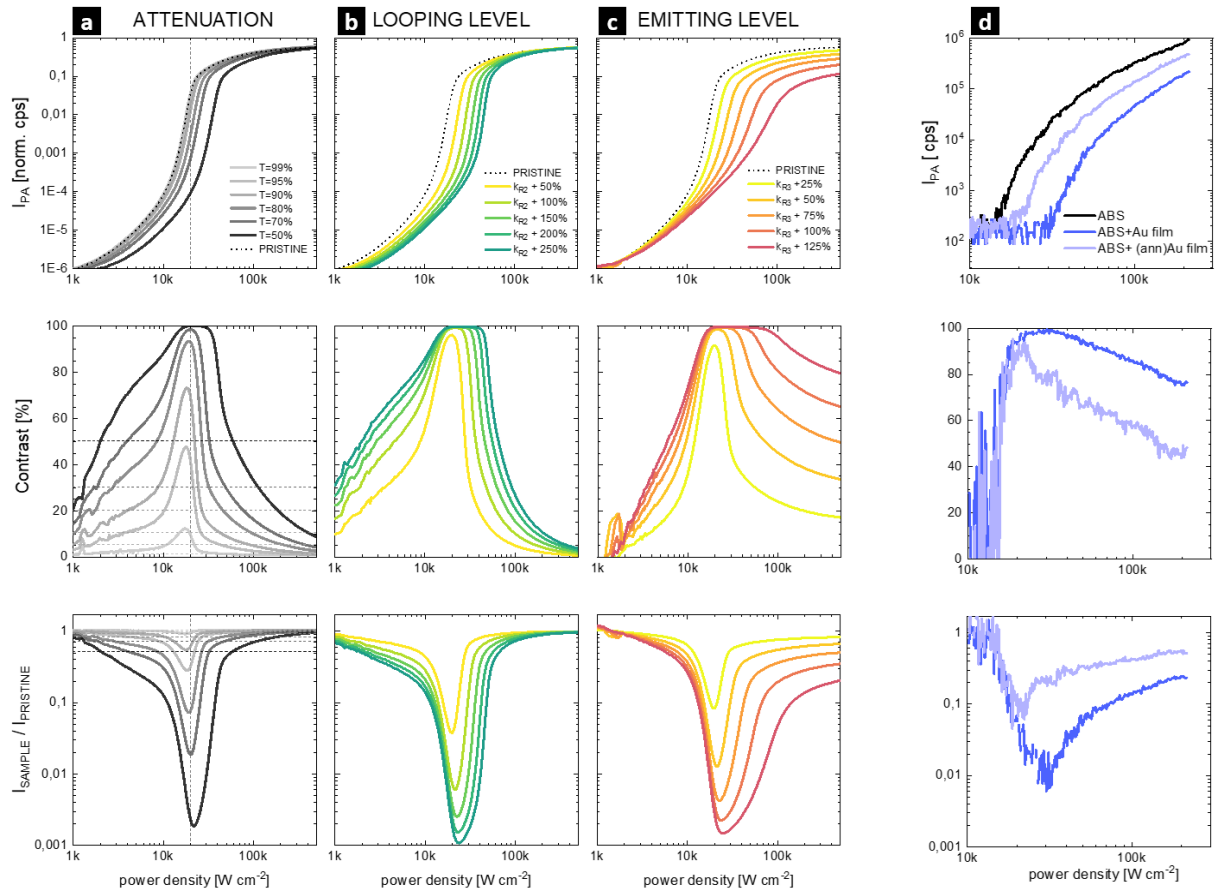

**Figure 3. Simulations of pump power dependent PA emission intensity of the ABS** in response to (a) pure attenuation of the excitation beam, (b) quenching of the looping level, (c) quenching of the emitting level – all plots were confronted with the pristine PA behaviour (dotted line on panels a-c). The variability is presented in response to percentage change of (a) transmittance (T=99, 95, 90, 80, 70, 50%) of the sample; (b) radiative rates of the looping level (+50, 100, 150, 200, 250% of $k_{R2}$) and (c) radiative rates of the emitting levels (+25%, 50, 70, 100, 125% of $k_{R3}$) for 3% Tm doping versus the pristine conditions and rates. The corresponding contrast (2$^{nd}$ row) and attenuation ($I_S/I_O$ – 3$^{rd}$ row) pump-power dependent profiles are presented against profiles obtained for the pristine ABS. For pure attenuation horizontal dashed lines indicate intrinsic attenuation level. Panel (d) shows how the ABS reacts to pristine and annealed gold film (discussed later), placed in the path of the excitation beam in TAB configuration.

Equipped with the fundamental knowledge on the ABS quality and homogeneity, and the precautions related to the interpretation of the PA signal, we performed TAB imaging of various objects. We started with thin gold film deposited directly on the thick glass microscopy slide (*Preparation of gold film* chapter). The ABS crystals were randomly dropped on the glass slide with Au film structures. In **Figure 4a**, the ABS was laying on the edge of the thin Au film. While the white light contrast does not exceed $C_{WL}$=31%, at 1064 nm, where the real interaction occurs, this contrast is equal to 40% (**Figure S3.2**). The loss of PA emission derived from TAB image (**Figure 4b - black line, Figure 4c)** offers substantially improved contrast equal to 90%. The cross-section through the edge was studied versus pump power Similar experiment was performed for homogenous Au film with a 3 um thick scratch (**Figure 4g-h**). Again, the contrast has been enhanced from original 21% to 96%. Next, we have exploited the laser induced annealing (LIA) of thin gold film exposed to strong and tightly focused laser beam, aiming to demonstrate *in-situ* modification and evaluation of this transformation. Macroscopically, upon LIA @ 1060 nm, gold nano-micro structures were formed – this was supported by the appearance of plasmonic peak at 550 nm (*Figure S4.1f*) and illustrated with SEM images (*Figure S4.5*). Because these metallic nanostructures does not feature optical activity corresponding neither to the excitation or emission bands of $Tm^{3+}$ ions, we expect no plasmonic interaction to be visualized in the course of TAB imaging. However, some changes of the S-shape profiles could be noticed when comparing ABS on glass, pristine and annealed Au film (*Figure 4e*). This conclusion is further supported by the contrast and LIR profiles (*Figure 4d*), which resemble quenching of the emitting level (*Figure 4c*) for the gold film, which became reduced for the annealed version of the gold sample. For the TAB experiments, a single ABS crystal was located at the edge of the fresh, thin Au film, and was imaged, both, in white light, as well as in TAB configuration (*Figure 4k*). Next laser annealing was performed in selected area of the sample using raster scanning with purposefully high 1.02 MW/cm² laser beam and imaged again in white light and in TAB configuration. The laser annealing leads to significant increase of transmission (*Figure S4.1*) at 1064 nm from 71.8% for pristine gold to 96.1 % for the annealed region. The TAB image demonstrates recovery of the PA intensity in the area, where ABS crystal was shielded by the Au film, which clearly indicates the capability to perform in situ observation of the Au film modification. In the same time, the ABS substrate is very stable in terms of photostability – it does not photo bleach nor photo blink, thus reliability of imaging is related purely to technical issues, such as long term power stability of the single mode laser source, and thermally stable conditions for the PA substrate [62]. Photodarkening has been reported in ANP, but our experiments excluded this effects under high pumping power and 120 s exposure, which did not modify the PA emission intensity in ABS (*Figure S3.2*). Concurrently, the PA substrate is simple to make with current, well established, reproducible and optimized hydrothermal synthesis protocols [54].

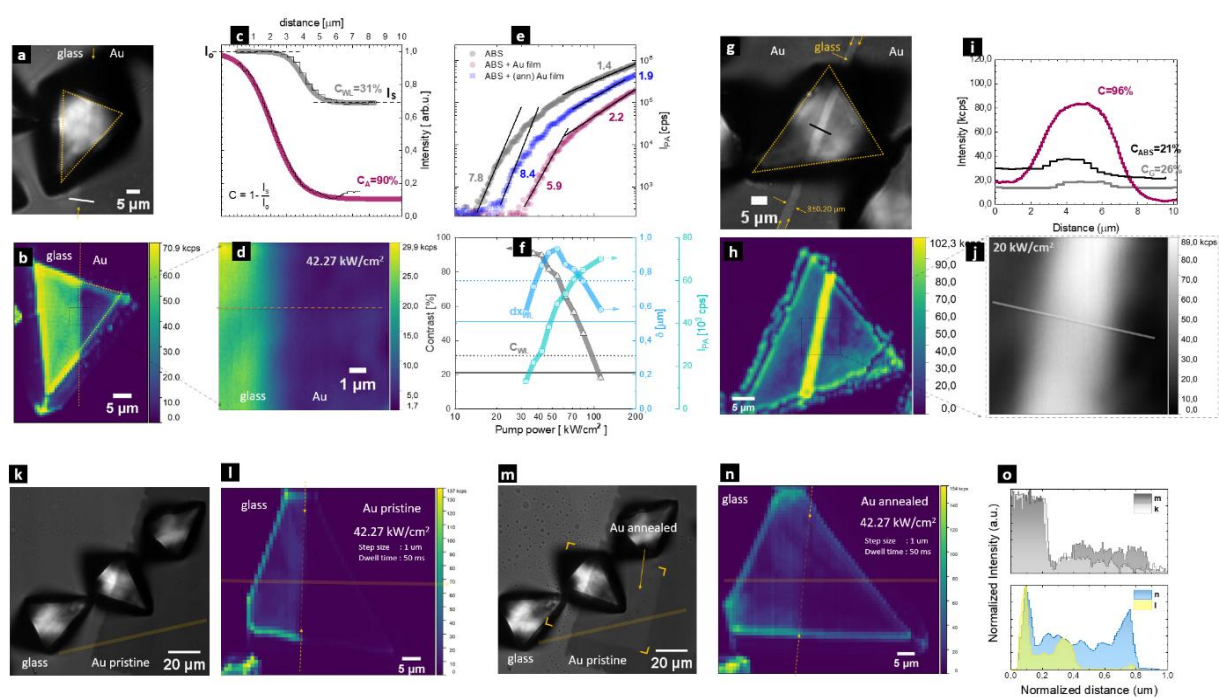

**Figure 4**. TAB Imaging of thin gold films. (a) White light and (b,d) the corresponding TAB image of the ABS laying across Au film. (c) cross section of intensity along the lines indicated on (a) and (b); (e) pump power dependent PA emission intensity for the bare ABS (black) and ABS through the fresh (purple) and annealed (blue) Au thin film; (f) corresponding (to (e)) pump power dependent contrast C (grey, left axis), intensity (green, right axis) and width (blue, right axis). (g-j) (g) White light and (h-j) TAB image of the Au film with a scratch, (i) cross sections through these (g,j) images along the lines;

(k-o) dynamic studies of Au film before (k-l) and after (m-n) laser annealing, (o) cross-section through the lines indicated on white light (m,k) and TAB (n,l) panels.

So far, we have evaluated the contrast augmentation on macroscopic samples, however, TAB imaging allows also to obtain enhanced optical resolution. To demonstrate this feature we performed local LIA of the Au Film and placed the ABS-es on such pretreated glass slide (**Figure S4.2-S4.4**). As a result, the gold naturally gathered into small particles to reduce its surface energy. This process of laser-induced annealing (LIA) caused the smooth film to turn into separated gold nanostructures. White light (WL) images barely indicate the presence of these LIA spots (**Figure S4.4a-c**), and the obtained images are of insufficient contrast nor does reveal details, even if software postprocessing (histogram correction) is applied (**Figure S4.4d-e**). In stark contrast to WL, TAB images reveal many interesting features (**Figure S4.4f-g**). First of all the contrast is much improved (LIR~150/90=1,6(6) for TAB vs 1.09 for WL for raw, unprocessed images). Secondly, the TAB image clearly and reproducibly shows doughnut shape, which based on SEM images (**Figure S4.5b-c**), clearly indicate the LIA process with Gaussian beam is not homogenous, namely, in the center of the beam produced Au islands that are larger and more sparsely distributed (case I), while the LIE with the wings of laser beam resulted in Au islands that are smaller and much closer to each other (case II). To clearly demonstrate the enhanced optical resolution we have performed LIA on much larger area of Au film using high power stimulation, aiming to generate case I islands, which are within the expected optical resolution range TAB that is resulting from S~10 in our ABS crystals. Indeed, SEM images confirmed formation of Au islands with average size equal 84.7 nm, and are displaced by average 126.7 nm (**Figure 5**, **Figure S4.6-S4.7**). The obtained TAB images showed clear pump power dependence of the resulting optical resolution, namely, under stronger excitation power (closer to saturation region of PA performance) the TAB images are brighter but more blurred as compared to these captured with lower laser photoexcitation (at the most steep part of pump-power curve), when less photons are collected but the features are more crisp in shape. While great majority of these Au islands are individual objects (**Figure 5a, ROI$_{1,3}$**), occasionally during LIA process, some Au islands are forming larger, irregular structures (**Figure 5a, ROI$_2$**). These two representative types of Au islands are TAB imaged (**Figure 5b,d**) showing distinct features and structures. The intensity cross-section profiles (**Figure 5c,e** correspondingly) clearly demonstrate features, i.e. dimming resulting from shielding the photoexcitation beam by the Au islands. We do not expect plasmonic effects to occur in this case because of two reasons – the large size of Au islands and excessive distance between the Au island and the ABS crystal, which is simply dropped on the previously prepared sample of gold islands. These features sizes are equal to 50-70 nm – slightly more than the sizes of Au islands and Au structures derived from the representative images (**Figure S4.6e**), and definitely below theoretical diffraction limit of light (<250 um, down to 108 nm for objective 100x NA=1.45 or c.a. and $\lambda$=1064 nm) with c.a. 4-fold improved optical resolution.

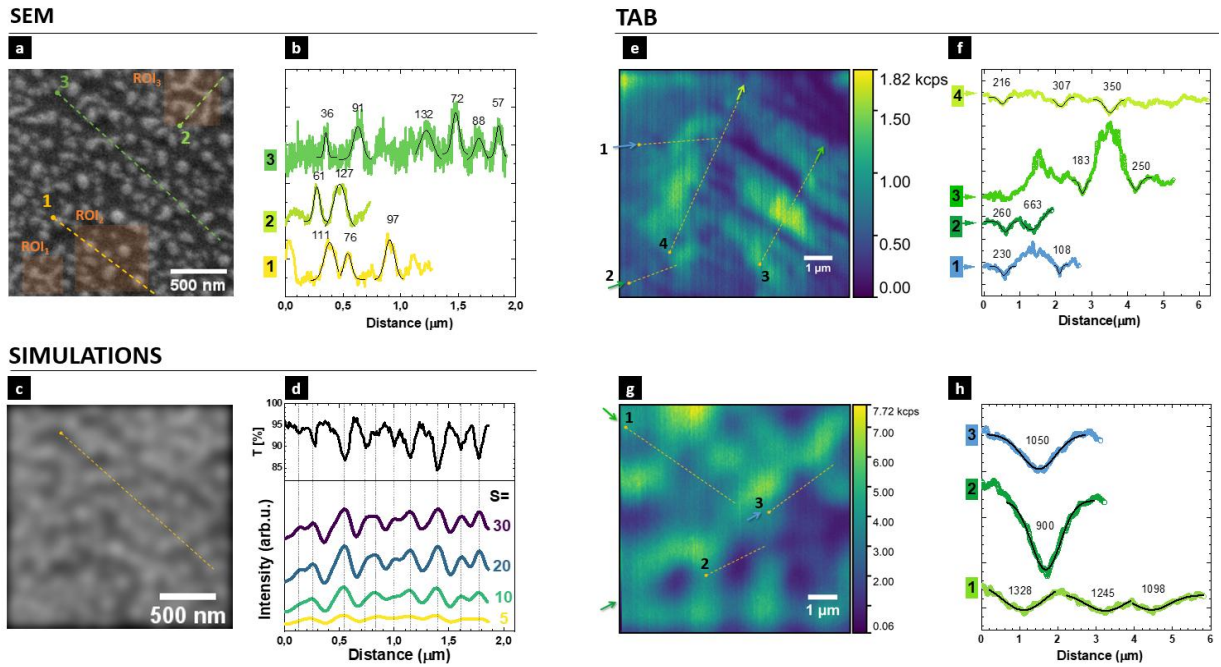

**Figure 5. Comparison of SEM, TAB Images and TAB image simulations of gold islands.** (a) SEM image of representative Au film after power laser treatment; ROI$_{1,2,3}$ indicate various types of Au islands (b) cross-sections along the solid lines indicated on panel (a); (c) simulations of TAB imaging using full SEM image (from a) as the sample (transmission 20%, non-linearity index S =10, NA=1.45, top panel of (d)); (d-bottom) cross sections of simulated TAB map – the sample (a, dashed line) and corresponding TAB intensity cross section extracted along yellow line on (c). Scale bars are 500 nm, both c) and d) show inverted intensities. (e, g) Representative experimental TAB images of Au islands; the heterogenous (e) and homogenous (g) regions (similarly to the one presented on panel (a) with red and orange frames, respectively ) are shown together with the corresponding cross sections (f) and (h) along the lines visible on panels (e) and (g). Intensity profiles correspond to TAB imaging mode – e.g. darker spots indicate obstacles (here Au islands), while brighter areas correspond to the ABS background emission. TAB maps collected with objectives (f) 100x NA=1.45 and (h) 40x NA=0.95. Dots on the lines (on a,c,e,g panels) indicate the beginning of cross-section line profile.

Being an all-optical, facile and affordable, light diffraction unlimited, the TAB imaging pave new way to study biological or photonic sub-100 nm objects with enhanced contrast. To enable further progress, here we critically evaluate some potential challenges to be addressed in the future studies. Frist of all, the TAB imaging operates in quasi far-field – the virtual aperture is generated in far field, but the actual imaging occurs mimics a near-field imaging principle. Therefore, the increased distance between the sample or obstacle from the ABS crystal will hamper the optical resolution unless Bessel beams are used. Moreover, the TAB imaging is not resolvable in axial direction, and the laser beam for TAB raster scanning may potentially interact with the sample. These effects will strongly depend on the pumping wavelength and power density – the kilowatts per square centimeter used currently are a value acceptable for many applications, however it also remains sufficient to potentially induce thermal, plasmonic, autofluorescence or optical trapping effects. For example, occasionally, the focused beam leaded to rotation and translation of our ABS crystals, therefore one of the important future challenges to be addressed is the development of large and thin avalanching ABS substrates. This can be achieved by a number of possible ways: (i) development of synthesis of nanosheet crystals [63], (ii) depositing avalanching thin shell on the non-luminescent (micro-bulk) crystals , (iii) self-assembly of avalanching nanocrystals to form a flat, highly uniform and dense monolayer of ANPs[64], or (iv) ions implantation into to surface bulk single crystal. To operate in TAB mode with the best available resolution and contrast, it is critical to use perfectly homogenous ABS crystals - free from nanoclusters of dopants, defects or cracks to assure perfectly homogenous backlight and avoid locally enhanced scattering and brightness variation. Moreover, development of various PA crystals working at various excitation wavelengths from UV-Vis-NIR regions would be beneficial to significantly extend the palette of the obstacles available for imaging, or even to gain more information about sample details and spectral characteristics due to differential measurements at various excitation lines. At the moment, emission at 475 and 800 nm under 1059 nm in $Tm^{3+}$ doped microcrystals have been observed. Moreover emission at 484, 525, 609, 642 nm (from $Pr^{3+}$), 450, 475, 650 and 800 nm (from $Tm^{3+}$) and 977 nm (from $Yb^{3+}$) have been observed under 852 nm photoexcitation. Other wavelength combinations are possible to achieve photon avalanche emission (**Figure S5**) [52,65,66] and further progress is expected. [50] This pose some potential new challenges - from one hand side shorter wavelengths assure smaller diffraction limited spot, smaller VA and higher optical resolution, from the other these shorter wavelengths may induce autofluorescence of the samples, which may increase background signal. Moreover, shorter wavelengths mean, that both, emitting and looping levels fall into the excitation bands of possible endogenous fluorophores, which may initiate resonant energy transfer or plasmonic interaction between the ABS and the sample. These interaction are strongly distance dependent. Moreover currently, due to the pyramidal, multi-facet shape of our ABS crystals, the only optical configuration was the inverted microscope configuration, where the ABS was laying on the sample. The up-right microscope configuration would be beneficial for number of reasons. In the latter case, the sample could be adhered to the surface of flat ABS crystal that would be imaged from the top – such approach could enable visualization of thicker biological samples, biomolecules, photonic devices and nanomaterials of various thickness.

Despite these predicted challenges, there are numerous advantages the TAB microscopy may offer. First of all, it is optical non-destructive far-field technique which allows to image samples with no pre-treatment, which are neither inherently fluorescent nor labeled. Taking into account perfect photostability of PA crystals and lack of blinking, as well as low toxicity of this material, long observation times are possible, enabling *in-situ* time-laps experiments. This technique may be suitable for imaging of the very transparent biological samples – for example the 1059 nm excitation and 800 nm emission [51,54] used here, are safe for biological samples [67] and do not generate background autofluorescence of the sample. The only requirement for the sample is, to be capable to absorb or scatter the excitation wavelength, which is used to induce PA emission in ABS crystals. To assure broader applicability, other PA wavelength combinations should be available at the other spectral ranges, and while not prerequisite, contrast agents may further augment the specificity of imaging. The interaction of the sample with the PA substrate in-situ, for example due to resonant energy transfer, plasmonic interactions or quenching of superficial layers of PA substrate [44] may further enhance the contrast of imaging but is not prerequisite. Nevertheless, studying pump power dependence (PPD) in these cases, may further reveal the complex nature of these interactions and will enable to distinguish pure attenuation from more complex energy transfer or quenching mechanisms, aiming to evaluate highly transparent (photonic, biological) samples or thin layers. Due to susceptibility of PA phenomenon to temperature, environment or applied force, such TAB method may be also suitable to visualize temperature or strain distribution maps or study bio-interactions at nanometre scale [58,59,61,62]. The large (anti)Stokes shift and NIR-Vis operation range allow to benefit from lack of sample autofluorescence, and thus also combination of TAB with conventional fluorescence methods (wide field, confocal or super-resolution) or near field techniques (SNOM, TERS, TENOM) is feasible and advantageous.

In conclusion, we proposed, devised and evaluated a novel concept of transmission avalanche backlight microscopy whose major advantages include label-free imaging with augmented imaging contrast and enhanced optical resolution. Not only macroscopic feature such as enhanced contrast of Au thin film was revealed, but TAB let us also study the microscopic scale formation of Au islands under focused laser beam. This approach allowed to visualise 30-50 nm large Au islands as 110-250 nm objects below diffraction limit of light. Moreover, TAB revealed homogenous and inhomogeneous regions occurring from interaction of gaussian laser beam with thin Au film. The presented TAB imaging approach may be suitable for enhanced, in-vitro and label-free imaging of native morphology, pristine subcellular architecture, revealing unperturbed dynamics and functioning of living cells, subcellular structures or photonic devices at nanoscale, without the risk of their irreversible damage or malfunctioning induced by excessive or exogenous labelling.

**MATERIALS AND METHODS**

*Contrast enhancement*

When the excitation beam passes through the attenuating medium of transmission $T[\%]$, then the original pumping power ($I_P$) becomes attenuated ($I'_P$), i.e. $I'_P = T \cdot I_P$. However, in PA regime, the luminescence intensity $I_L$ is proportional to the power S of pumping intensity $I_P$, where S is the non-linearity parameter in the power law: $I_L = (I_P)^S$. Therefore, the PA luminescence intensity equals $I'_L = (T \cdot I_P)^S = T^S \cdot I_L$, i.e. for T<1, it is lower by a factor of $T^S$ as compared to the original, nonaffected luminescence intensity $I_L$ (**Figure 1b, d-f**). In consequence, any non-zero attenuation of this pumping light beam will be translated to disproportionally lower photon avalanche signal. The luminescence intensity loss can be quantified by luminescence intensity ratio (LIR) between the PA luminescence of the beam attenuated by the sample, related to the PA emission intensity of the non-attenuated beam equals

$$LIR = \frac{I'_L}{I_L} = \frac{(T \cdot I_P)^S}{(I_P)^S} = T^S \qquad \text{Eq.1.}$$

where the non-linearity S is pump power dependent ($S(I_P)$). To experimentally quantify the performance of TAB substrate, a contrast $C(I_P)$ parameter can be defined as

$$C(I_P) = \frac{I_L - I'_L}{I_L} = \left(1 - T^{S(I_P)}\right) \qquad \text{Eq.2.}$$

and the absolute sensitivity

$$S_A(I_P) = 100\% \frac{\delta I_L}{\delta T} = 100\% \frac{\frac{I_L - I'_L}{I_L}}{1 - T} = 100\% \frac{1 - T^{S(I_P)}}{1 - T} \qquad \text{Eq.3.}$$

defines what relative change in PA emission intensity occurs in response to relative attenuation change. In other words, $C(I_P)$ measures how much the TAB augments the visualization of highly transparent subjects as compared to white light observation. For example, assuming photon avalanche nonlinearity of S=20, the pristine attenuation by 1% ($T = 99\%$), will be evidenced by dimming PA luminescence from the incident one $(I_P)^S$ down to on $(T \cdot I_P)^S$, which is equal to $I'_L = T^S \cdot I_L = (0.99)^{20} \cdot I_L = 0.8179 \cdot I_L$. This leads to translation of 1% change to luminescence intensity decrease by 18.2%, thus 8.2-fold better contrast. Moreover, the transmission $T(\lambda)$ is related to attenuation $A_T(\lambda)$ by a simple formula $T(\lambda) = 10^{-A_T(\lambda)}$ and indirectly, through Beer-Lambert Law $A_T[M^{-1}cm^{-1}] = \varepsilon c l$, to the molar attenuation coefficient $\varepsilon[cm^{-1}]$, molar concentration $c[M^{-1}]$ of these species as well as the pathlength $l[cm]$. With an appropriate calibration, TAB imaging can therefore enable quantitative measurements of absorption, concentration or thickness of thin layers.

*TAB simulations*

The Matlab 2019b simulations carried on Fig.2 assumed a diffraction limited excitation ($\lambda_P$=1059nm) gaussian beam is raster scanned through the phantom (grey layers or green particles) whose transmission $T(x)$ is defined as pink area on Fig.2.. The PA backlight substrate is therefore illuminated with a pump profile $I_P(x) = I_{PO} \cdot \exp(-\frac{(x-x_0)^2}{\delta_{p0}^2})$, where $x_0$ and $\delta_0$ are central position and full width half maximum ($\delta_{p0} = \frac{\lambda_p}{2 \cdot NA}$) of the excitation spot respectively. This initial pump beam profile is multiplied by the transmission of the phantom $T(x)$, thus the effective pump intensity profile is defined as $I'_P(x) = T(x) \cdot I_P(x)$. For linear substrate the luminescence intensity is proportional to the pump intensity, but for photon avalanche case the luminescence scales with power S. Moreover, every illuminated spot along the x axis generates a diffraction limited luminescence ($\lambda_L$=800nm) gaussian beam $\delta_{L0} = \frac{\lambda_L}{2 \cdot NA}$, which after summing these contribution up $I_L(x_0) = \sum_x \left[ T(x) \cdot I_{P0} \cdot \exp(-\frac{(x-x_0)^2}{\delta_{p0}^2}) \right]^S$ is ascribed as luminescence intensity at position $x_0$. The signal is then normalized to the luminescence intensity achieved for the non-perturbed case (for $T(x) = 1$). For simulation presented on Fig.5c, the same method was applied in 2D. The phantom was based on real SEM image, i.e. the 8-bit gray scale SEM image $I_{SEM}(x,y)$ was scaled to get the phantom image $I_{ph}(x,y)$, as $I_{ph}(x,y)$= 1.0-(T/100)*($I_{SEM}(x,y)$/255) with modulation depth 20%.

*Modeling of PA properties*

The TAB method is cable to not only enhance the contrast, but can also evidence some physical interactions between ABS substrate and the sample. When the PPD measured for the sample resembles that of pristine ABS (with the only difference the PA threshold being higher for the sample), the TAB image occurs exclusively due to attenuation or scattering of the sample and no physical interaction shall be expected. However, as soon as the PPD of the sample differs from the PDD of the pristine ABS, quenching of enhancement must be taken into account.

The modeling of PA behavior was based on the numerically solved differential rate equation presented previously, which were modified aiming to, on-by-one investigate the impact of attenuation (transmission $T = 1-A$ = 50, 70, 80, 90, 95 and 99%), quenching of emitting (non-radiative rate increase by $dk_2$ = 50, 100, 150, 200 and 250% of the emitting level radiative rate) and looping (non-radiative rate increase by $dk_3$ = 25, 50, 75, 100 and 125% of the looping level radiative rate) levels (presented in Figure 3) and impact of cross relaxation ($dk_{CR}$ = -20, -10, -5, 5, 10, 20% change versus the original CR rates). The impact of those factors was investigated by substituting $I'_P = (1-T) \cdot I_P$ (for attenuation), $A'_3 = (1 + \delta k_3) \cdot A_3$ (for emitting level quenching), $A'_2 = (1 + \delta k_2) \cdot A_2$ (for looping level quenching) and $f_{CR} = (1 + \delta k_{CR})$ (for discussion on the homogeneity of doping across large PA crystals)

$$\frac{dn_3}{dt} = \frac{I_P}{h\nu} f_{ESA}\sigma_{ESA}n_2 - (A_3 + k_3)n_3 - f_{CR}s_{31}n_1n_3 + f_{CR}Q_{23}n_2^2 \qquad \text{Eq.1}$$

$$\frac{dn_2}{dt} = \frac{I_P}{h\nu}(\sigma_{GSA}n_1 - f_{ESA}\sigma_{ESA}n_2) - (A_2 + k_2)n_2 + (\beta_{32}A_3 + k_3)n_3 \\ + 2f_{s31}s_{31}n_1n_3 - (f_{CR}Q_{22} + 2f_{CR}Q_{23})n_2^2 \qquad \text{Eq.2}$$

$$\frac{dn_1}{dt} = -\frac{dn_2}{dt} - \frac{dn_3}{dt} \qquad \text{Eq.3}$$

where $\sum_{i=1..3} \frac{dn_i}{dt} = 1$ and $\sum_{i=1..3} n_i = 1$. The $I_P$, $\sigma_{GSA}$, $\sigma_{ESA}$ parameters denote pump intensity, GSA and ESA cross section, $s_{31}$ denotes the CR rate between level 3 and 1, $k_3$ and $k_2$ are non-radiative multi-phonon relaxation rates and $\beta_{ab}$ denotes the emission branching ratio between state $n_3$ and $n_2/n_1$ ($\beta_{32}$ = 0.144; $\beta_{31}$ = 1 − $\beta_{32}$ = 0.856); $A_a$ is the radiative rate from level $a$ ($a$=2,3). Following,[68] the other parameters are equal to $s_{31} = a_{cr} \cdot c^2$ ($a_{cr}$ = 160s$^{-1}$, c=8%), $Q_{22} = a_{uc} \cdot \frac{c^3}{c^2+4.3^2}$ ($a_{uc}$ = 9.00s$^{-1}$), $Q_{23} = a_{inv} \cdot \frac{c^3}{c^2+4.3^2}$ ($a_{inv}$ = 25.6s$^{-1}$). $A_2$ = 162 s$^{-1}$ and $A_3$ = 636 s$^{-1}$.

In the steady-state conditions (i.e., when $\frac{dn_i}{dt} = 0$ occurs for all levels), the population of the emitting level, $n_3$, and thus, the relation between excitation and emission intensity ($A_3 n_3$) can be derived. These simulations are aiming to provide physical context and educative hints on the susceptibility of PA to the attenuation of the excitation beam, the non-radiative depopulation of the emitting ($n_3$) and looping ($n_2$) levels las well as the impact of the CR rates, rather than deliver exact solutions and fit our experimental data.

*Reagents*

Microcrystals were prepared using commercially available reagents. Yttrium oxide $Y_2O_3$ (99.999%), thulium oxide $Tm_2O_3$ (99.995%) were purchased from Chemicals101Corp, ammonium fluoride $NH_4F$ (≥98.0%) was purchased from Sigma-Aldrich. Ethanol (96% pure p.a.) and nitric acid $HNO_3$ (65% pure) were purchased from POCH S.A. Ethylenediaminetetraacetic acid EDTA (≥99%) and lithium fluoride LiF (99%) were purchased from POL-AURA.

*Synthesis of 3%Tm LiYF$_4$ microcrystals*

LiYF$_4$:Tm$^{3+}$ microcrystals were synthesized by reported earlier hydrothermal synthesis method with some modifications [https://doi.org/10.1021/acsami.0c07765]. 20 mmol of RE(NO$_3$)$_3$ (Y(NO$_3$)$_3$ and Tm(NO$_3$)$_3$) and 1.06 g EDTA were dissolved in 30 mL distilled water. The mixture was vigorously stirred at room temperature for one hour (solution 1). Simultaneously, 20 mmol of LiF and 60 mmol of NH$_4$F were mixed with 30 mL of distilled water (solution 2). Next, solution 1 was added dropwise to the solution 2 under vigorous stirring. The white thick mixture was then stirred for 30 minutes and transferred to a 100 mL Teflon-lined autoclave. The synthesis was carried out for 12 hours at 200°C. The product was then allowed to cool naturally to room temperature. White precipitate was separated from supernatant and washed multiple times with distilled water and ethanol using an ultrasonic bath. The obtained microcrystals were dried on a hotplate at 100°C for 2 hours. The structure and morphology of Avalanche Backlight Substrate (ABS) were presented on Figure S2.1.

*SEM-EDS measurements for 3%Tm LiYF$_4$ microcrystals*

The morphology and chemical composition of Tm$^{3+}$-doped LiYF$_4$ microcrystals were investigated by using a FE-SEM FEI NovaNano SEM 230 (FEI Company as a part of Thermo Fisher Scientific) equipped with an energy dispersive X-ray spectrometer EDAX Genesis XM4 with the resolution better than 135 eV and compatible with Genesis EDAX Microanalysis Software. Prior to the implementation of SEM imaging, the microcrystal sample was dispersed in alcohol and subsequently deposited onto a carbon stub in order to eliminate charging and drift problems. SEM images were recorded by using secondary electron (SE) at an accelerating voltage equal to 5.0 kV. The chemical composition analysis was conducted using the EDS technique for a randomly selected LiYF4:Tm$^{3+}$ microcrystal. EDS measurements were acquired at an accelerating voltage of 30 kV using drift mode to enhance EDS maps quality. However, due to the inherent limitations of the EDS method in terms of the detection of light elements (Z ≤ 4), the EDS map of the Li element was not included and discussed in the paper.

*Preparation of gold film*

The substrate was cleaned by sonicating it in acetone, Isopropyl Alcohol (IPA), and water for 5 minutes. The substrate was then blow-dried using N$_2$ gas. A dehydration bake was performed at 120 $^0$C for 15 minutes. A patterned Au film with a thickness of 5 nm was deposited on the cleaned substrate using a thermal evaporator via the shadow masking technique at a vacuum of 7 X 10$^{-6}$ torr.

*SEM-EDS measurements for Au films*

The morphology of Au films were investigated by using the same equipment as for the Tm$^{3+}$-doped LiYF4 microcrystals i.e. FE-SEM FEI NovaNano SEM 230 equipped with an energy dispersive X-ray spectrometer EDAX Genesis XM4. Initially, the samples were positioned on the aluminum stub to rectify issues pertaining to charging effect. Subsequently, the samples were placed under the microscope to record SEM images at an accelerating voltage of 2.0 kV in a beam deceleration mode, with the objective of obtaining more detailed features of the samples.

*Photophysical PA properties evaluation and TAB imaging*

Spectroscopic characterization of the samples was carried out using a home-built fluorescence microscope based on the body of Nikon Ti-U Eclipse inverted microscope and shown schematically in **Figure S6.1**. Excitation light at 1059 nm was provided by a single mode laser module (1064CHP, TEM$_{00}$ Gaussian mode, 3SP Technologies), and controlled and thermally stabilized by a dedicated controller (ITC4001 controller + LM14S2 butterfly mount, Thorlabs). In order to automatically and precisely regulate the power of produced beam, motorized system of neutral density filters (NDF) was utilized. First of them (NDF1) was a gradient neutral density filter (NDC-100C-4M, Thorlabs) used for gradual tuning of output power, while the other (NDF2) set of neutral density filters (FW121CNEB with preloaded ND filters, Thorlabs) was used for rough selection of the range of accessible light powers. After passing this module the beam was directed to the glass slide reflecting part of radiation to the optical power meter sensor (S121C+PM100, Thorlabs) and reference power value is monitored, which can be further used to determine the power density of the light illuminating the sample. Most of light (90%) is transmitted through the beam splitter. To overfill the back aperture of the objective, the laser beam is expanded using manual beam expander. Subsequently, the beam is directed by a dichroic mirror (DM2, DMSP900R, Thorlabs) to the lens (L) and enters the microscope's side port. Inside the microscope, light is reflected by the mirror (M) and, after passing through the lenses (L), enters the objective lens (O, Plan Apo λ 40x, NA=0.95, Nikon, and oil immersion Plan Apo λ 100x, NA=1.45, Nikon), where it is focused on the sample. The sample stage is integrated with an xyz piezoelectric stage (PI-545 system, and E-727.3CDA controller, Physik Instrumente) for raster scanning. The same objective lens is utilized for collecting of the sample emission and directing it back through the system of mirrors (M) and Lenses (L). Next, the beam is transmitted by the DM2 and passes through a set of filter wheels consisting of neutral density filters (ND3 = OD1.0, OD2.0, or OD4.0, Thorlabs), which are utilized to limit the intensity of signals exceed the dynamic range of the detectors. Once formed and pre-filtered, the emission signal can be directed through two equal focal length bio-convex lenses and a 200 μm pinhole (P200K, Thorlabs), which is kept at the focal point. The signal then passes through a band-pass filter (BPF, FB800-40 Thorlabs) before entering the avalanche photodiode unit (APD Counts© Laser Components). The photon counts are integrated with a PCIe-6351 National Instruments PCI card, which are synchronized with piezo stage movements. Alternatively, emission light can be directed to the setup dedicated to emission spectra measurements, composed of Andor Shamrock 500i Czerny-Turner type monochromator with 3 gratings turret (GT) and equipped with the Andor Newton 920P-BEX2-DD CCD camera with active area chip cooled down to -60°C to limit the detector dark current. White light images were acquired with Andor iXion 897 512x512 EMCCD camera.

*Acknowledgements*

The authors acknowledge financial support from NCN, Poland, grant number NCN 2021/43/B/ST5/01244. MSz acknowledge financial support from NCN, Poland, grant number NCN 2024/55/D/ST11/02790. The authors wish to thank dr Dawid Piątkowski (UAM, Toruń, Poland) and M.Sc. Paweł Szczypkowski (UW, Warsaw, Poland) for discussion and technical support during development of the TAB instrument. The authors thank dr Bagusław Macalik for absorption measurmeents.

**SUPLEMENTARY MATERIALS**

Supplementary material for this article is available